# Extrinsic Dopants as Growth Modifiers in Cu-Cr-O delafossites: A Study of Incorporation Limits and Film Properties

Marco Moreira[1,2], Yves Fleming[1], Patrick Grysan[1], Christele Vergne[1], Adrian Marie Philippe[1], Petru Lunca-Popa* [1,2]

[1]Luxembourg Institute of Science and Technology, 41 Rue du Brill, L-4422 Belvaux, Luxembourg
[2]University of Luxembourg, 2 Avenue de l'Universite L, 4365 Esch-sur-Alzette

*Corresponding author. petru.luncapopa@list.lu

**Abstract**

Cu–Cr–O delafossite thin films were grown by metal–organic chemical vapor deposition with various extrinsic dopants (Al, Mg, Mn, Sc, Y, and Zn) targeted at 5 at. % to investigate how such doping influences their structure and properties. X-ray photoelectron spectroscopy revealed that the actual dopant incorporation is well below the nominal 5%, with only Al and Sc present above detection. An off-stoichiometric $CuCrO_{2+0.15}$ composition is determined, with no secondary phases detected. Transmission electron microscopy indicates that films grown on c-plane sapphire are epitaxial near the substrate interface but relax into a polycrystalline structure beyond ~40 nm, while films on silicon are polycrystalline throughout. All films show high p-type conductivity (on the order of $10\text{-}10^2$ S·cm$^{-1}$) attributable to the excess oxygen, with no significant variation among different dopants. Optical transmission measurements indicate a slight red-shift (~20 nm) of the absorption edge for all doped films, likely arising from strain effects and subtle structural disorder introduced during growth. We discuss the influence of lattice strain (investigated by X-ray diffraction sin²ψ measurements showing residual strain) and small-polaron absorption behavior in these films. Despite limited incorporation of dopants, subtle structural and optical shifts suggest that dopant precursor chemistry and growth conditions play a significant role in influencing film stoichiometry and properties.

**Introduction**

Transparent conducting oxides (TCOs) are a cornerstone of modern optoelectronic devices, enabling applications ranging from flat-panel displays and touchscreens to solar cells, LEDs, and transparent electronics. While n-type TCOs are well-established in commercial technology—most notably indium tin oxide (ITO)—the reliance on scarce and expensive indium has prompted intense research into more sustainable alternatives. Materials such as Al-doped ZnO (AZO), Ga-doped ZnO (GZO), and Sn-doped $In_2O_3$ (ITO variants) have been explored to balance performance with abundance and cost.[1, 2] In contrast, the development of p-type TCOs remains far more challenging due to the inherent difficulty in achieving simultaneous optical transparency and sufficient hole conductivity [3–6] . Among the various systems investigated, delafossite oxides, especially Cu-based compounds like $CuCrO_2$, have emerged as leading candidates due to their layered structure and hybridized Cu3d - O2p valence band[7, 8], which offers better hole transport pathways. However, delafossites face significant limitations, including low hole mobility and doping constraints, which hinder their practical deployment in high-performance devices.[9] Nonetheless, not all applications demand high mobility. Devices such as gas sensors [10–12], rectifiers, photodetectors[13, 14] buffer layers in photovoltaics [15–17], and n-type junction field-effect transistors can still exploit the unique attributes of delafossites and other p-type transparent conducting materials, without being limited by mobility constraints. Their chemical stability and earth-abundant constituents make delafossites attractive for scalable device integration. In its ideal stoichiometry, $CuCrO_2$ adopts a rhombohedral structure (space group $R\bar{3}m$), consisting of layers of linearly coordinated $Cu^+$ ions stacked between sheets of edge-sharing $CrO_6$ octahedra. Delafossites are appealing due to their relatively simple scalable [15] deposition processes and wide optical band gaps; however, their performance is fundamentally constrained by deep acceptor levels and the resulting polaronic conductivity mechanisms[18]. In its undoped form, $CuCrO_2$ is a poor semiconductor. Slight off-stoichiometry—such as copper deficiency or oxygen incorporation[19]—is often intentionally introduced to generate hole carriers and improve conductivity by several orders of magnitude compared to the stoichiometric compound.[20–23] Another strategy to tailor the optoelectronic properties of $CuCrO_2$ is extrinsic cation doping[24, 25]. Prior studies have shown that substituting a small fraction of Cr sites with aliovalent dopants can introduce additional holes or modify the band structure. For instance, $Mg^{2+}$ substitution at the $Cu^+$ site has been demonstrated to enhance p-type conductivity significantly in Cu-based delafossites.[26–28] Despite these promising strategies, extrinsic doping of delafossite oxides remains challenging. The rigid delafossite lattice exhibits low dopant solubility and often responds to substitution by forming secondary phases (e.g., spinel oxides or CuO) or by excluding the dopant from incorporation altogether [8, 19, 29]. This can result in actual dopant concentrations being far lower than intended, accompanied by unintended defects or lattice strain.

Moreover, in many studies, the reported compositions correspond to the nominal or target stoichiometry rather than the actual composition of the synthesized material. For example, in sol-gel methods, the cited ratios typically reflect precursor concentrations used during synthesis, without subsequent verification of the final product's composition. Similarly, in sputtering, the target composition is commonly reported, while the deposited film may exhibit deviations due to preferential sputtering or re-sputtering. Even in pulsed laser deposition (PLD), which is generally regarded as a stoichiometry-preserving method, discrepancies can arise due to elemental volatility, plume dynamics, or substrate-induced effects. These examples underscore a broader methodological concern: the need for rigorous post-synthesis compositional analysis when correlating material properties with chemical composition. It should be emphasized that this observation is not intended as a critique of any specific study or author, but rather as a constructive point to improve experimental reliability and reproducibility in the field. Another critical factor in extrinsic doping is the potential for lattice distortion and induced strain. Beyond the generation of charge carriers—which directly affects conduction mechanisms—dopants can alter lattice parameters. The interplay between extrinsic dopants, defect chemistry, and strain may significantly influence the electronic structure and optoelectronic properties of $CuCrO_2$. Studies have shown that tuning the hole effective mass in polaronic materials is inherently difficult. While strain can impact carrier mobility, it typically fails to elevate it to levels required for high-performance applications. [30]

In this work, we investigate extrinsic doping of Cu–Cr–O delafossite thin films with a series of dopant cations ($Al^{3+}$, $Mg^{2+}$, $Mn^{3+/2+}$, $Sc^{3+}$, $Y^{3+}$, and $Zn^{2+}$), introduced during MOCVD growth. The choice of dopants was guided by a combination of factors including ionic radius, valence state, and reported behavior in related delafossite systems. These elements span a range of ionic radii from relatively small ($Al^{3+}$ ~0.535 Å) to large ($Y^{3+}$ ~0.9 Å in octahedral coordination), allowing assessment of how lattice distortion and strain correlate with incorporation and film properties. Their valence states also differ: $Mg^{2+}$ and $Zn^{2+}$ introduce a lower charge state than $Cr^{3+}$, potentially creating acceptor-like behavior or compensating defects; $Al^{3+}$, $Sc^{3+}$, and $Y^{3+}$ are isovalent but differ in size and chemical compatibility. Mn, with multiple possible oxidation states ($Mn^{2+}/Mn^{3+}/Mn^{4+}$), was selected to explore redox-sensitive incorporation pathways. Rather than focusing solely on improving conductivity, our study aims to elucidate dopant solubility within the $CuCrO_2$ lattice, the structural distortions and strain induced by doping, and the consequent effects on optical absorption—particularly near the band edge—and electrical transport. By integrating compositional analysis with structural, microstructural, optical, and electrical characterization, we provide a comprehensive understanding of how selected extrinsic dopants influence $CuCrO_2$ thin films. Although extrinsic doping is commonly pursued to enhance electrical conductivity or optical transparency in delafossite oxides, practical doping is frequently hindered by solubility limitations and chemical compatibility issues. In this study, we deliberately explored these boundaries using identical β-diketonate (THD) precursors for all metals (Cu, Cr, and dopants), allowing a direct comparison of their intrinsic incorporation barriers. Even in cases where minimal dopant incorporation occurred, the observed subtle but consistent changes in structural strain and optical absorption underline that dopants may exert significant influence through surface or vapor-phase interactions rather than bulk substitution alone. These insights may support future efforts in tailoring delafossite oxides through controlled doping and defect engineering.

**Material and Methods**

**Thin Film Growth.** Thin films were deposited on $Al_2O_3$ c-cut and on Si (100) substrates using a Dynamic Liquid Injection. Metal Organic Chemical Vapor Deposition system (DLI-MOCVD, MC200 from Annealsys). Metal based THD (2, 2, 6, 6-tetramethyl-3, 5-heptanedionate) precursors were used in all cases. Cyclohexane was used as a solvent. A 2.5 mM equimolar Cu/Cr solution was used in the case of undoped solution. All precursor were mixed in a same injection cannister. To introduce extrinsic dopants (Al (III), Mg (II), Mn (III), Sc (III) Y (III) and Zn (II)), appropriate metal-organic precursors were added to the solution at concentrations corresponding to 5 atomic% of the total cation content (Cu + Cr + dopant). The deposition parameters were previous optimized for maximum conductivity of delafossite: [15, 31, 32] temperature substrate = 450 °C; oxygen/nitrogen atmosphere, total process pressure = 12 mbar. The deposition time was set to 60 minutes, yielding films with a thickness of ~300 nm.

**Chemical and Structural Characterization.** The chemical composition and depth profiles of the films were analyzed by X-ray photoelectron spectroscopy (XPS). The elemental composition of the doped samples was investigated by X-ray photoelectron spectroscopy (XPS) on a Thermo Scientific Nexsa G2 spectrometer equipped with a monochromatic Al Kα X-ray source (hν =1486.7 eV) operating at 60 W. A sequential 2kV Ar+ ion beam was used to etch the films layers by layers on a 1 mm x 2 mm area. High-resolution spectra were acquired using a 50-eV pass energy and a 200 μm spot size. The compositional depth profiles were then determined with the CasaXPS software (version 2.3.26) using the sensitivity factors provided by the instrument manufacturer. Grazing-incidence X-ray diffraction (GI-XRD) was employed to identify the crystalline phases present in the films. The patterns were recorded on a Bruker D8 diffractometer (Cu Kα radiation) at a shallow incidence angle to maximize film signal. High-resolution θ–2θ scans were acquired at various tilt angles (ψ) of the sample ($\sin^2\psi$ method) to assess lattice spacing changes under different orientations. The (006) and (012) reflections of the delafossite phase were tracked as a function of ψ.

Transmission electron microscopy (TEM) was used to examine the film microstructure and the film/substrate interface. Cross-sectional TEM specimens were prepared with a FEI Nanolab 650 Focused Ion Beam - Scanning Electron Microscope (FIB-SEM) following the usual "lift-out" procedure. Crystallographic identification and orientation study was conducted by performing Fast Fourier Transform (FFT) analysis of High-resolution TEM (HRTEM) micrographs acquired with JEOL F200 cold-FEG microscope operated at 200 kV. Further analysis of the crystalline nature of the thin films was achieved through Selected-Area Electron Diffraction (SAED) measurements. Scanning TEM (STEM) combined with energy dispersive X-ray spectroscopy (EDS) mapping was performed to investigate the spatial distribution of Cu, Cr, O, and dopant elements within the films.

Surface morphology was characterized by atomic force microscopy (AFM). (MFP-3D Infinity, Oxford Instruments). The scans were collected in tapping mode on a 5×5 μm^2 area of the film surface to determine the surface roughness and grain size. The root-mean-square (RMS) roughness was extracted, and grain size distributions were estimated (e.g., by a watershed algorithm) from the AFM topography.

**Optical and Electrical Measurements.** Optical reflectance and transmittance spectra of the films (on sapphire substrates) were recorded using a UV–Vis–NIR PerkinElmer 1050 spectrophotometer in the wavelength range 250–2500 nm. The transmittance (T) data were converted to absorption coefficient (α) spectra using the relation $\alpha = -(1/d)\cdot \ln T$, where $d$ is the film thickness. From the calculated absorption spectra, Tauc plots were constructed to estimate the direct optical band gap (the linear extrapolation of the absorption edge to α=0). For electrical characterization, the sheet resistance of each film (on insulating substrates) was measured at room temperature by a standard four-point probe method. Sheet resistance values were converted to bulk conductivity (σ) using the film thickness determined from TEM/AFM.

**Results**

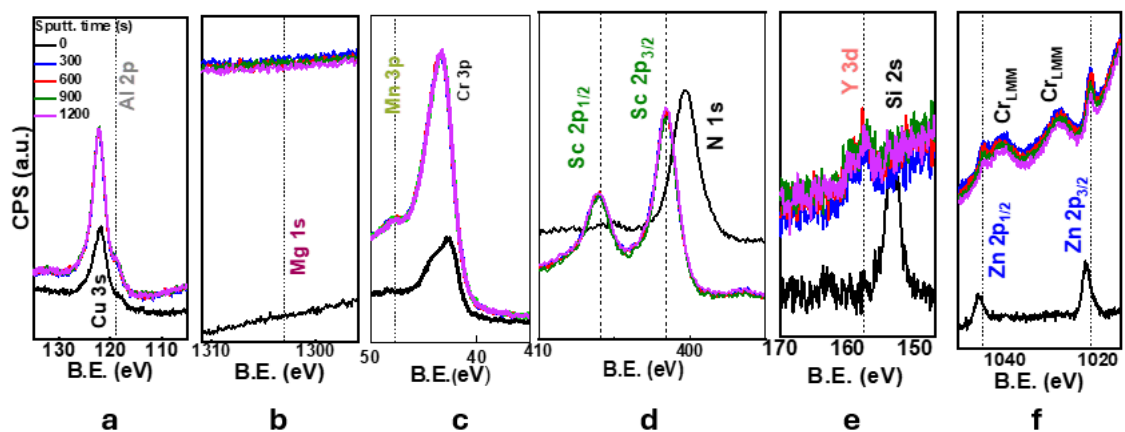

**Figure 1.** XPS spectra in the binding energy range of (a) Al 2p, (b) Mg 1s, (c) Mn 3p, (d) Sc 2p, (e) Y 3d and (f) Zn 2p.

**Chemical Composition and Dopant Incorporation**

XPS analysis provided quantitative information on the elemental composition of the as-deposited doped Cu–Cr–O films. The initial surface analysis indicated the presence of Cu, Cr, O, some of the respective dopant elements (Al, Mn, Sc, Y, Zn) as well as carbon, silicon, calcium and nitrogen. After sputtering off a few nanometers, the extraneous signals of adventitious (C, Si, Ca, N) disappeared, indicating that those elements are only superficial contaminants, likely due to exposure to air and handling, and are not incorporated in the bulk Cu-Cr-O films. Figure 1 shows high-resolution XPS spectra in the characteristic binding energy regions of each dopant element (e.g. Al 2p, Mg 1s, Mn 3p, Sc 2p, Y 3d and Zn 2p). For manganese, the secondary Mn 3p peak has been scanned due to the overlap between Mn 2p and Cu LMM peaks. All narrow scans were collected at the surface (sputtering time = 0 s) and at varying depths. These in-depth spectra confirm that the intended dopants are present in the film matrix to some extent, except Mg which is neither detected at the surface nor in the bulk film. The 5 other dopant signals (Al 2p, Mn 3p, Sc 2p, Y 3d and Zn 2p) were all detected in the bulk but at different levels as described next.

Figure 2 depicts the XPS depth profile quantification, showing the atomic concentration (at.%) of each element as a function of sputtering time for the six different dopant cases. The key finding is that the actual incorporated dopant concentrations are far lower than the 5 at. % nominal value introduced in the precursor. Among the six dopants, only aluminum and scandium were present in measurable quantities in the bulk of the films: Al at about 1.5 at. % and Sc at about 2.2 at. % (after surface contamination). In contrast, Mn, Y, and Zn were detected only at trace levels close to 0.5 at. % detection limit of our XPS system. Magnesium could not be detected at all in the bulk region of the Mg-doped film. None of the doped samples approached the targeted 5 at. % dopant level.

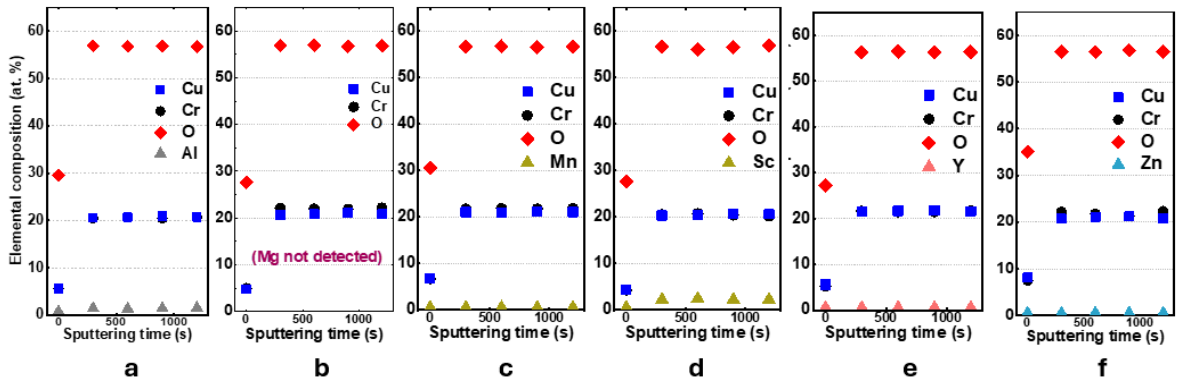

**Figure 2.** In depth elemental composition depth profiles of Cu–Cr–O as deposited films doped with (a) Al, (b) Mg, (c) Mn, (d) Sc, (e) Y and (f) Zn (targeted concentration 5%)

An important observation is that the Cu:Cr ratio in these films is approximately 1:1. Cu and Cr each constitute roughly 20–22 at. % of the film, rather than 25 at. % (which would correspond to exactly $CuCrO_2$). Oxygen, on the other hand, is found to be around ~57 at. % (instead of 50 at. % for ideal stoichiometry). This indicates that the films are oxygen-rich, with an approximate formula $CuCrO_2+\delta$ where $\delta \approx 0.15$ (i.e., about 15% more oxygen than the stoichiometric amount.

**Microstructure Analysis**

Two representative samples were chosen for detailed TEM analysis: one doped with Sc grown on a sapphire substrate, and one doped with Y) grown on a silicon substrate (both 5 at. % nominal. These two cases allow a comparison of a sample with measurable dopant incorporation (Sc ~2.2 at. %) versus one with minimal dopant incorporation (Y <0.5 at. %), as well as the influence of different substrates. Figure 3a shows a cross-sectional high-resolution TEM micrograph of the Sc-doped Cu–Cr–O film on α-Al$_2$O$_3$. The film exhibits two distinct regions: near the film–substrate interface (within the first 10-20 nm of the film), the lattice image indicates an epitaxial relationship with the substrate. The Fast Fourier Transform (FFT) from this interfacial region, can be indexed consistently with the sapphire substrate diffraction (notably the 006 reflection of CuCrO$_2$ aligning with the substrate lattice). However, beyond a certain thickness (~20 nm), the film's microstructure changes: the lattice becomes more disordered and grain boundaries appear, suggesting a transition from epitaxial to a polycrystalline growth mode. This behavior is common for heteroepitaxial films [33] when a critical thickness is exceeded: strain energy or slight lattice mismatch between CuCrO$_2$ and sapphire likely leads to the formation of misfit dislocations or grain boundaries, causing the upper layers to recrystallize in random orientations to relieve stress.

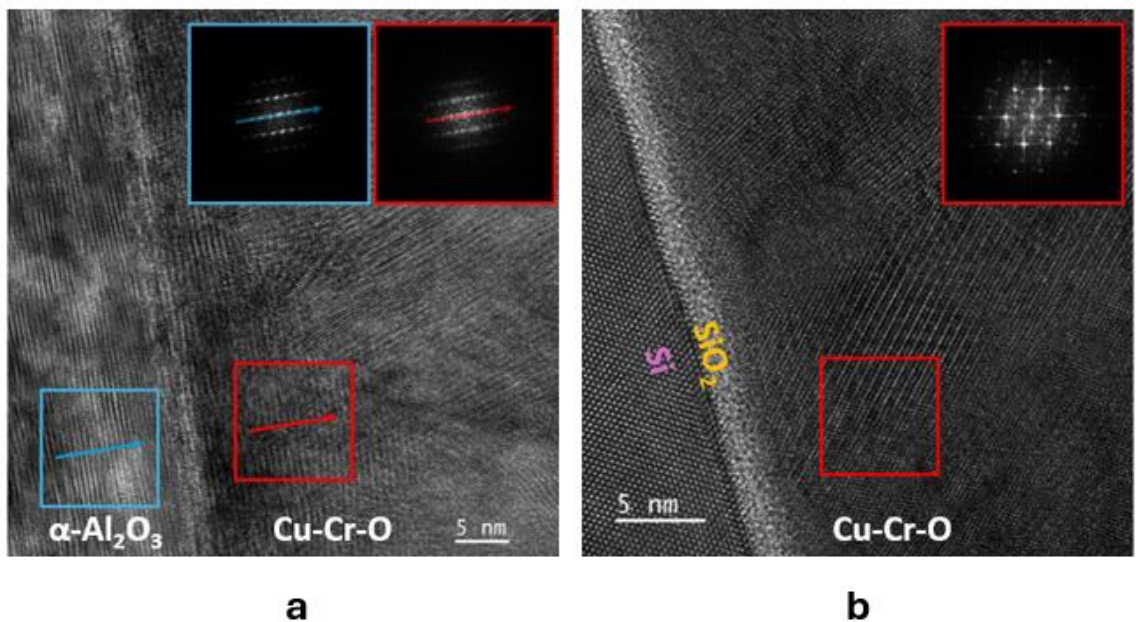

**Figure 3.** HRTEM micrographs of Cu–Cr–O films doped with: Sc on an α-$Al_2O_3$ substrate, and (b) on a Si substrate. The insets show the FFT patterns from the film and substrate regions

Figure 3b depicts a cross-sectional HRTEM image of the Y-doped Cu–Cr–O film on a Si substrate. As expected in this case, the film is polycrystalline from the very interface, since the CuCrO$_2$ delafossite structure has no epitaxial registry with the amorphous native oxide and dissimilar lattice of silicon. The FFT indicates that the diffraction pattern of the film consists of spots from multiple orientations, characteristic of random polycrystalline structure. Thus, the Y-doped film on Si grows in a polycrystalline manner throughout its thickness. Secondary phase regions or precipitates in either the Sc- or Y-doped TEM samples were not observed. The film thickness was also estimated. The Sc-doped film has an average thickness of 300 nm whilst the Y-doped one measured 200 nm. As the deposition parameters were identical, this discrepancy suggests that the introduction of the Y precursor altered the growth kinetics or yield. (for example, by introducing gas-phase species that partially inhibited the deposition).

Figure 4 presents the STEM-EDS dopant elemental maps for the Sc-doped sample on sapphire. Overall, the STEM analysis demonstrates that the films are chemically homogeneous on the nanoscale, with no evidence of phase separation. The Sc map (Figure 4a) indicates that Sc is homogenously incorporated in the film, consistent with the XPS measurement.

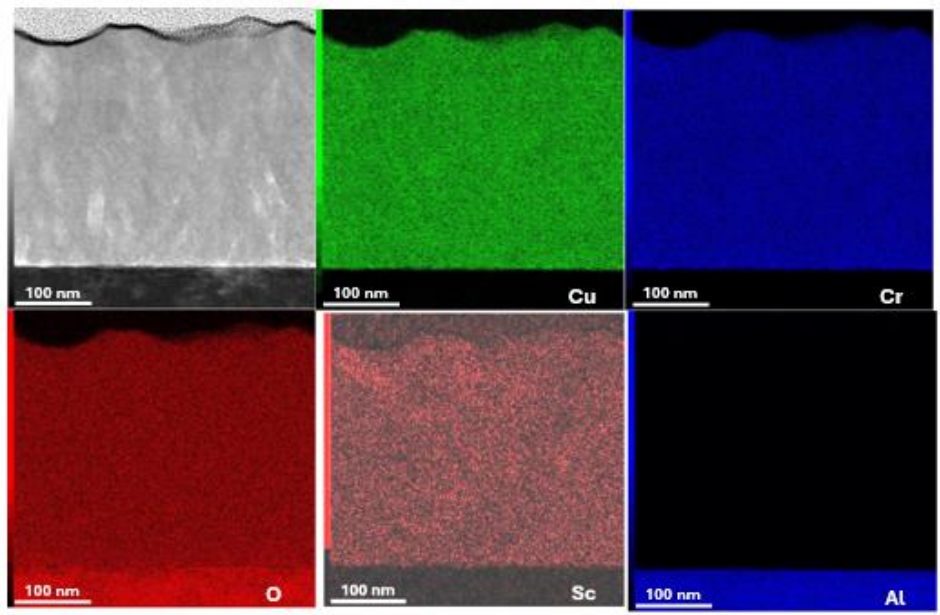

**Figure 4**. STEM-EDS analysis of a cross-section of the Cu–Cr–O film doped with Sc on α-Al$_2$O$_3$ substrate.

The STEM analysis for CuCrO$_2$:Y demonstrates also a uniform distribution of Cu, Cr and O but with no Y being detected in the film (the Y content is too low to be detected by EDS-X).

**Surface Topography.** Figure 5 compares the 3D surface topography of two same samples as in the TEM section: Sc and Y -doped CuCrO$_2$. Both films exhibit a similar nanoscale morphology characterized by closely packed grains. The surface consists of nanometric grains giving a somewhat rough texture by visual inspection. Quantitatively, the root-mean-square (RMS) surface roughness was measured to be around 11.6 nm for the Sc-doped film and 10.3 nm for the Y-doped film. The grain size distribution extracted from the AFM images (using watershed segmentation to delineate individual surface grains) also shows similar mean grain diameters for the two samples: approximately 71 nm ± 24 nm for CuCrO$_2$:Sc and 76 nm ± 27 nm for CuCrO$_2$:Y, suggesting that the presence of the dopant at these low levels does not strongly affect the lateral grain growth. The similarity in surface morphology between Sc-doped and Y-doped films (despite differences in thickness and dopant incorporation) implies that the growth mode and surface evolution of the films are dictated more by the overall deposition conditions than by the specific dopant chemistry, at least for the low dopant levels realized. An approximative 10 nm roughness was measured, a level that might have implications for optical scattering and the accuracy of optical measurements, but relatively small compared to the film thickness.

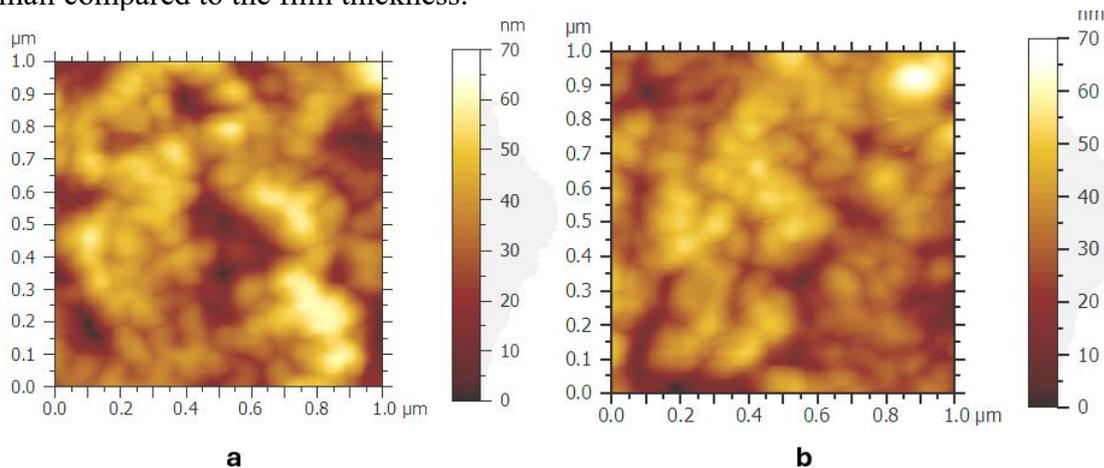

**Figure 5.** AFM surface topography of Cu–Cr–O films doped with: a - Sc and b – Y

**Optical Properties**

Reflection and transmission spectrum were acquired and the absorption and furthermore the absorption coefficient were estimated. Figure 6 presents the absorption coefficient (α) spectra for a representative set of films over the wavelength range 2500 nm to 250 nm (corresponding to photon energies from ~0.5 eV to ~5 eV). The most notable feature in these spectra is a shift in the fundamental absorption edge between the undoped and doped samples. The undoped $CuCrO_2$ film shows an absorption onset (rapid increase in α) near ~380 nm (which corresponds to about 3.26 eV, typical for $CuCrO_2$ band gap in slightly off-stoichiometric form). In contrast, all the doped films exhibit an approximative 20 nm redshift.

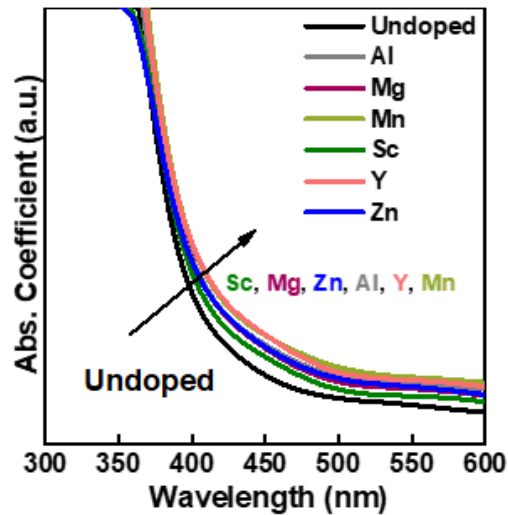

**Figure 6.** Absorption coefficient (α) spectra for undoped and doped Cu–Cr–O thin films (Al, Mg, Mn, Sc, Y, Zn dopants)

The usual method to determine the band gap in semiconductors is to use a Tauc plot, which assumes a particular functional form of α vs photon energy (e.g., $\alpha h\nu \propto (h\nu - E_g)^n$ for direct or indirect transitions with index $n$ depending on the nature of the transition). For most reports on $CuCrO_2$ (treated as a direct-gap semiconductor), $n = 2$ is used, and the linear portion of $(\alpha h\nu)^2$ vs hν is extrapolated to zero absorption to find Eg (direct allowed band gap). We have applied the same approach here.

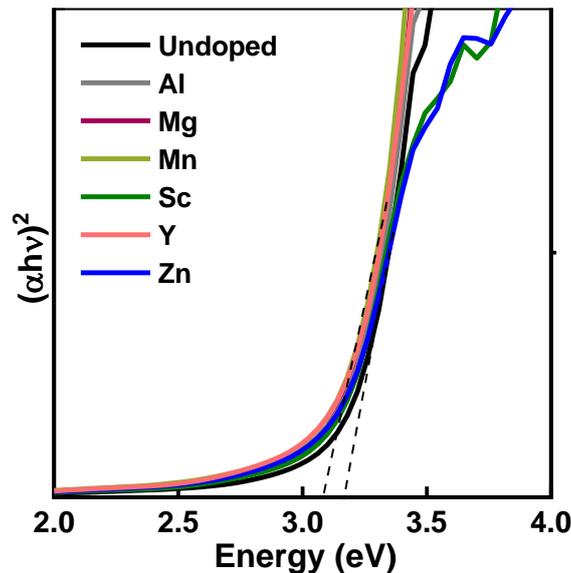

**Figure 7**. Tauc plots of undoped and doped Cu–Cr–O films

All the doped samples exhibit a marginally lower band gap in the Tauc plot. The reduction is small but systematic: for each doped film, the extrapolated band gap is 0.05–0.1 eV lower than the undoped value. No obvious trend among different dopants, all leading to a similar outcome for the optical edge. This suggests that the mechanism for band gap reduction is not a conventional chemical or doping-specific effects.

**Crystal Structure and Strain (XRD Analysis)**

Figure 8 presents GI-XRD diffractograms for the series of films doped with Al, Mg, Mn, Sc, Y, and Zn (all on α-$Al_2O_3$ substrates). All major peaks in these patterns can be indexed to the $CuCrO_2$ delafossite phase, matching the reference pattern for $CuCrO_2$ with no extra peaks belonging to secondary phases. Specifically, peaks corresponding to the (012), (104), (006), (110), and other allowed reflections of the rhombohedral $CuCrO_2$ structure are observed. The GI-XRD mode emphasizes the film peaks over the substrate, but we note that there is a weak feature at ~41.7° 2θ position in some patterns that might correspond to some artefacts in this machine configuration. Importantly, all doped samples show the same set of $CuCrO_2$ peaks at roughly same positions, suggesting that doping did not drastically change the lattice parameters (which is expected given the low doping levels actually achieved). We do not observe any obvious systematic ordering of peak positions with dopant type, reinforcing that the doping concentrations are too low to measurably alter the lattice.

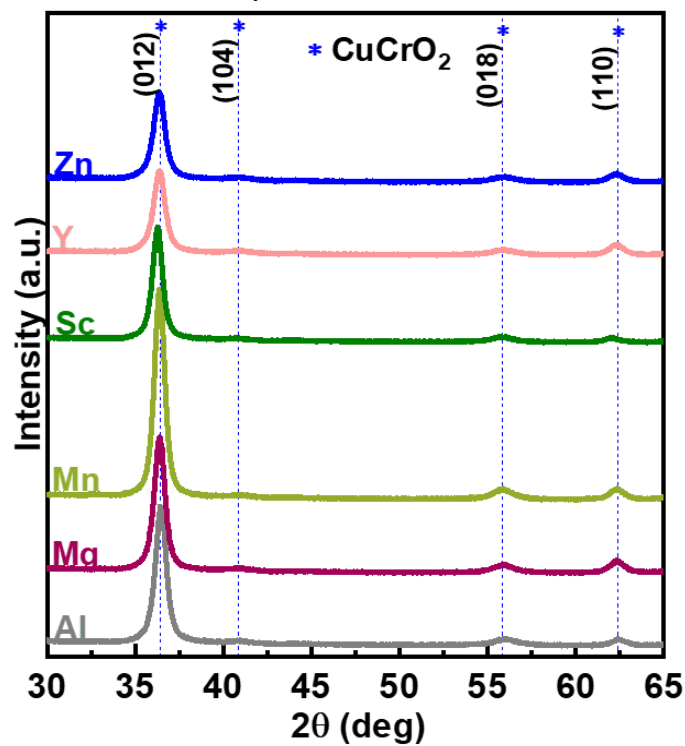

**Figure 8.** GI-XRD diffractograms of as-deposited Cu–Cr–O doped with Al, Mg, Mn, Sc, Y, and Zn (5% nominal each) on α-$Al_2O_3$ substrates. Curves are vertically offset for clarity.

To probe the strain state of the films, symmetric θ–2θ scans were conducted at multiple tilt angles (ψ) for each doped sample. Figure 9 depicts the scans for the doped films on sapphire substrates, focusing on the (006) and (012) reflections of $CuCrO_2$. The data for each dopant are plotted in panels (a) through (f). The angle ψ = 0° corresponds to the conventional Bragg-Brentano geometry (the diffraction plane is parallel to the film surface), which primarily measures out-of-plane lattice spacing (for the 006 reflection, ψ = 0 picks up the out-of-plane c-axis lattice spacing). As ψ is increased (tilting the sample such that we probe lattice planes inclined with respect to the surface), the diffraction condition for a given hkl reflects a combination of out-of-plane and in-plane lattice parameters.

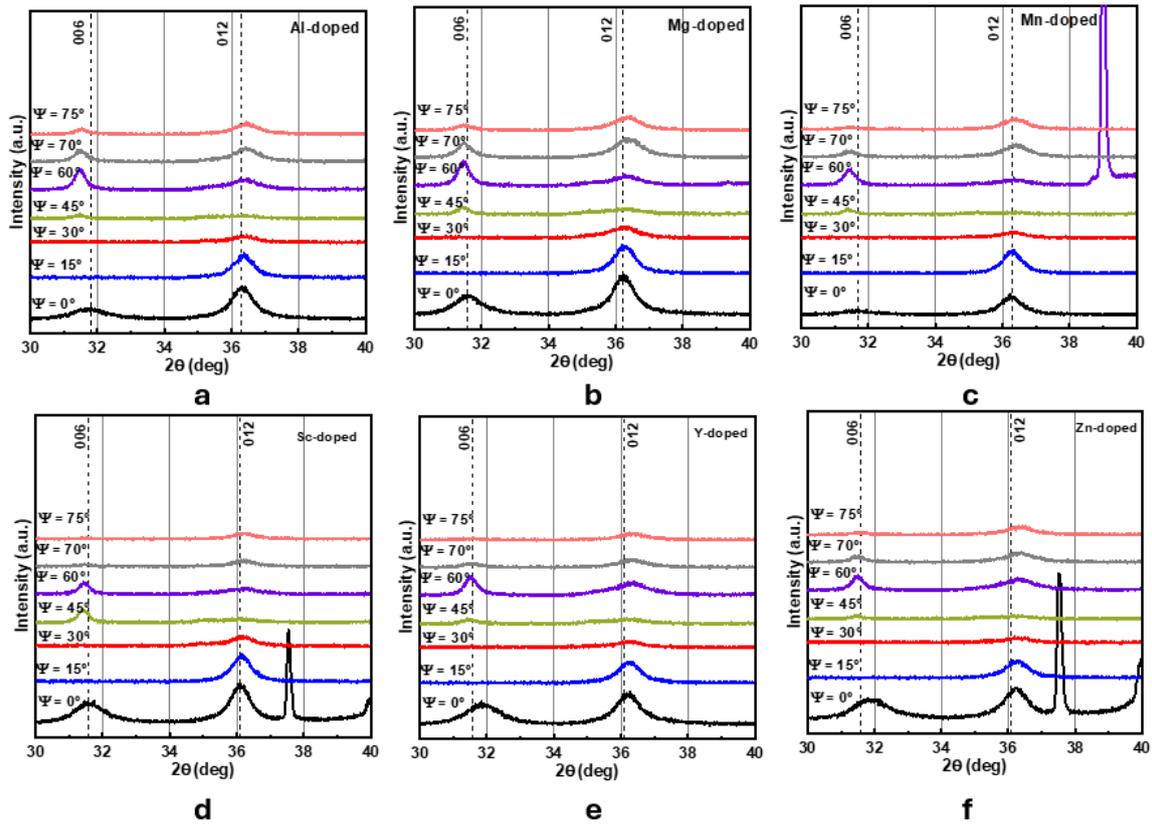

**Figure 9**. θ–2θ XRD patterns for doped Cu–Cr–O films with a - Al; b – Mg; c – Mn; d – Sc; e -Y and f - Zn on sapphire substrates α-Al$_2$O$_3$, collected at various tilt angles ψ

**Electrical Properties**

The electrical conductivity of the undoped and doped Cu–Cr–O films was measured to evaluate whether extrinsic dopants had any effect on charge transport. Figure 10 presents a comparison of the room-temperature electrical conductivities for all samples (undoped, Al-, Mg-, Mn-, Sc-, Y-, and Zn-doped), each grown on an insulating α-Al$_2$O$_3$ substrate. The conductivities were derived from four-point probe sheet resistance measurements, using the film thickness values from TEM for accuracy. All the films exhibit p-type conductivity values on the order of tens of S cm$^{-1}$. The "undoped" CuCrO$_2$ film shows a relatively high conductivity in this range, which can be attributed to its off-stoichiometry (δ ~ 0.15 oxygen excess, as determined by XPS). In stoichiometric CuCrO$_2$, conductivities are typically much lower (around $10^{-2}$ S cm$^{-1}$.

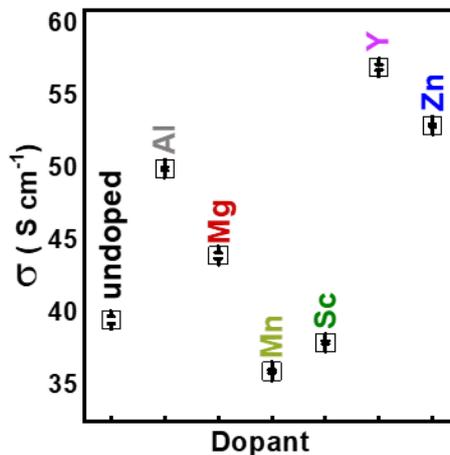

**Figure 10.** Electrical conductivity of non-doped and doped Cu-Cr-O samples on α-Al2O3 substrates.

None of the dopants produced a statistically significant change in conductivity. Within experimental uncertainty (given by example by the error from thickness non-uniformity), all doped samples have conductivity in the same range as the undoped sample.. If anything, one could say extrinsic doping did not improve conductivity beyond the baseline set by the intrinsic defects. This outcome is not unexpected considering the low actual dopant incorporation. As discussed, Mg and Zn (which in principle could act as acceptor dopants by substituting $Cu^+$ and providing holes) were not significantly present in the lattice, so they could not contribute additional holes. The fact that all films have similar conductivity suggests that the hole concentration is primarily governed by the off-stoichiometry given by Cu vacancies or interstitial Oxygen.

**Discussion**
In this study, all metalorganic precursors - Cu, Cr, and the extrinsic dopants (Al, Mg, Mn, Sc, Y, Zn)—were based on the same chelating ligand: 2,2,6,6-tetramethyl-3,5-heptanedione (THD). This is a β-diketonate ligand (example for the CuTHD molecule is depicted in figure 11) similar in structure to acetylacetonate (acac), but with bulkier alkyl substituents that increase thermal stability and volatility. The resulting metal–THD complexes ($M(THD)_3$ or $M(THD)_2$ depending on valency) are widely used in chemical vapor deposition due to their relatively high vapor pressure, clean decomposition, and compatibility with oxygen-containing environments. Using the same THD ligand for all precursors minimizes variability in evaporation rates, transport efficiency, and surface adsorption/reaction kinetics during MOCVD. This consistency helps ensure that observed differences in dopant incorporation or film properties arise from the metal centre's intrinsic chemistry (e.g., solubility, ionic radius, redox behavior), rather than ligand-related artifacts. The decomposition temperatures can vary based on the metal center but they all are under the actual process temperature. [34, 35].

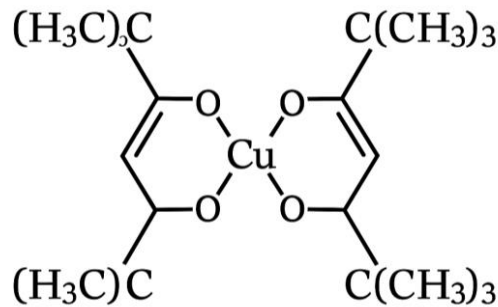

**Figure 11**. Schematic structure of a $CuTHD_2$ molecule

All dopants introduced at 5 at. % nominal ended up incorporating at the order of 0–2 at. % at most. This aligns with prior reports on bulk $CuCrO_2$ where dopant solubility is low and excess dopant often precipitates as separate phases (e.g., $MgCr_2O_4$ spinel or $Cr_2O_3$) rather than entering the lattice beyond ~1%[36]. The absence of secondary phase signatures in XRD or TEM suggests that excess dopant might have been lost during growth (e.g., evaporated or not deposited) rather than forming detectable precipitates. This indicates a strong tendency of the $CuCrO_2$ structure to resist extrinsic substitution beyond a quite low threshold. Factors contributing to this could include charge imbalance, size mismatch, and chemical incompatibility of the dopant cations.

While all doped films clearly differ from the undoped case, they do not show large differences among each other. Even dopants like Mg or Y (which were not detected in XPS) show essentially the same absorption edge shift as dopants like Sc or Al (which were detected at ~1–2 at.% in the film). This suggests that the mere presence of the dopant precursor during growth

– or the resultant slight changes in the film (whether composition, strain, or defect structure) – is sufficient to cause the optical shift, rather than the specific chemistry of the dopant. In other words, the effect is general for "doped vs undoped" but not highly sensitive to which dopant or how much was actually in the lattice.

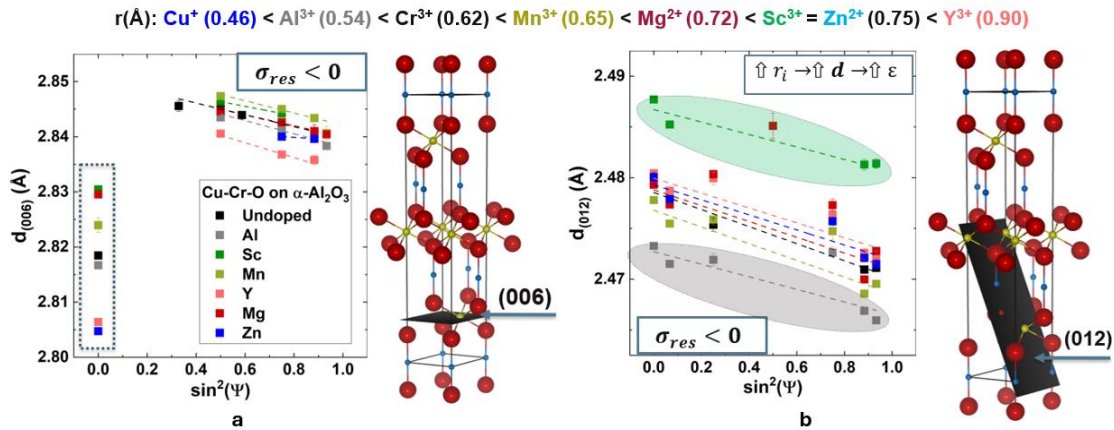

**Figure 12.** Calculated $\sin^2\Psi$ plots for a - 006 and b - 012 reflections of doped samples on sapphire substrates. The orientation off (006) and (012) diffraction planes are also depicted

**Strain analysis.** $\sin^2\psi$ method was used for investigating the strain in the films. Figure 12 plots the lattice spacing $d$ of the (006) reflection (panel a) and the (012) reflections (panel b) as a function of $\sin^2\psi$ for the doped films. A horizontal line indicates the absence of stress. Negative slopes were determined in the case of both (006) and (012) direction, indicating an in-plane compressive stress. Two main sources of strain are to be considered. The first one is related to the thermal mismatch between film and substrate. The films were grown at 400 °C and the measurements were performed at room temperature. For $CuCrO_2$ on $Al_2O_3$, literature suggests that $CuCrO_2$ has a higher thermal expansion than $Al_2O_3$,[30] that will induce a tensile strain. However, what we observe is compressive strain, which implies that another factor is dominating. This is very likely to be the lattice mismatch at the growth temperature (epitaxial constraint). As $CuCrO_2$ nucleates, it tries to align epitaxially to match the substrate spacing, (observed on TEM analysis), after which misfit dislocations release some strain. Nonetheless, some residual compressive strain remains in the film. The observation that the strain state appears to be compressive and the linear behavior suggests that the epitaxial (lattice mismatch) strain overshadowed the thermal strain. All samples showed a similar slope, implying comparable stress. A quantitative analysis on the relationship between the dopant and the strain in the film is very difficult to perform. First the lack of precise elastic stiffness values for $CuCrO_2$ impedes any exact numerical estimation. There are very scarce reports here, mostly for bulk stoichiometric delafossite. Nevertheless, we are able to identify some effects of the ionic radius. For the d plot of (006) reflection, all curves are close to that corresponding to the undoped sample. In typical R-3m stacking, this plane hits the midpoint between atomic layers, intersecting planes with maximum oxygen density and minimum cation density. The plots are more dispersed in the case of (012) reflection. It is clear that the film containing Sc (2.2%, r=0.75 A) shows larger $d_{012}$ parameter than that containing Al (1.5%, r=0.45A). For the samples with very low amount of dopant or with intermediate ionic radius the d plots are grouped around the undoped curve. These results suggest that the doping is done on the 012 plane whilst the 006 one remains quite unaffected by the doping process.

**Optoelectronic characterization.** Having established the limited dopant incorporation from XPS and the presence of compressive strain from XRD analysis, we now turn to electrical characterization to evaluate how these structural and chemical features influence the optoelectronic properties. Even very low doping levels—on the order of 0.1–2 at.%—can be considered significant in complex oxides, particularly given their typically narrow solubility ranges and sensitivity to local lattice distortion[36]. However, for the $CuCrO_2$ based films examined here, no notable differences were observed in the p-type electrical conductivity between doped and undoped samples. Conductivity values ranged from 35 to 70 S/cm across all samples, and no clear correlation with dopant type, nominal charge, or incorporation level could be established. It is important to clarify that the reference "undoped" $CuCrO_2$ film in this study is not strictly stoichiometric, but already intrinsically doped through oxygen non-stoichiometry. Pure, stoichiometric $CuCrO_2$ is known to be a poor conductor, and significant conductivity arises only in oxygen-rich or extrinsically doped variants[24]. In our case, all films—regardless of dopant—exhibit an appreciable excess of oxygen, with XPS and structural data suggesting $\delta \approx 0.15$. This oxygen excess likely contributes p-type carriers via the formation of $Cu^{2+}$ states or oxygen interstitials.

Comparable behavior has been reported in analogous systems such as $CuFeO_2$, where oxidation can lead to $CuFeO_2+\delta$ phases with $\delta$ up to ~0.18 under appropriate conditions.[37]. For smaller B-site cations with ionic radius below ~0.70 Å, the crystallographic site available for interstitial O is too confined, making such intercalation difficult without significant strain or defects [38]. Interestingly, our films show a consistent $\delta$ ~0.15 across different dopants, independent of B-site ionic radius. This suggests that the deposition conditions alone - rather than cation chemistry - govern oxygen uptake. Consequently, the observed electrical conductivity, ranging from 35 to 70 S/cm across all samples, appears to be primarily controlled by intrinsic defect chemistry, not by extrinsic dopant incorporation. No systematic correlation between conductivity and dopant type or nominal charge was identified, further underscoring the dominant role of oxygen-driven self-doping in these films. Extrinsic doping in such a case might serve more as a perturbation to the system rather than a primary driver of carrier generation.

For the undoped $CuCrO_2$ film, Tauc plot analysis yielded a direct optical band gap of approximately 3.20 eV, consistent with values reported for slightly off-stoichiometric $CuCrO_2$. Fully stoichiometric $CuCrO_2$, by contrast, typically exhibits a larger band gap in the range of ~3.3–3.4 eV. The reduction observed here is attributed to excess oxygen ($\delta > 0$), which introduces defect states or band tailing that effectively lower the apparent optical transition threshold. However, interpreting this "optical band gap" requires caution due to the small polaron conduction mechanism inherent in $CuCrO_2$.[39] In such materials, optical transitions may involve not only direct band-to-band excitations but also excitations from self-trapped (polaronic) states, which absorb photons at energies slightly below the nominal gap. Additionally, the exponent *n* in the Tauc relation may deviate from the ideal value of 2, since the transitions in polaronic systems do not strictly follow the selection rules for parabolic bands. Across all extrinsically doped films, a small but consistent reduction in the estimated optical band gap was observed relative to the undoped sample. Notably, this shift does not correlate strongly with dopant valence or identity, suggesting that the observed effect arises generically from the presence of extrinsic species during growth—likely mediated by strain, local disorder, or enhanced defect formation—rather than from direct electronic contributions of the dopants themselves. This reduction gap thus seems generic to the introduction of extrinsic impurities under our growth conditions.

Band gap narrowing in semiconductors can arise from several mechanisms beyond the classical Burstein–Moss (BM) effect. In n-type materials, BM typically causes a blue shift due to conduction band filling, while in p-type systems, such shifts are minimal owing to the higher density of states near the valence band maximum. In contrast, red shifts in the optical band gap—like those observed in our doped $CuCrO_2$ films—are more plausibly attributed to disorder-induced band tailing, localized defect states, or polaronic interactions. These mechanisms distort the electronic structure near the band edges, effectively reducing the apparent optical transition energy. Another possible explanation often discussed in degenerately doped semiconductors is band gap renormalization, a many-body effect where high carrier concentrations screen Coulomb interactions, leading to a narrowing of the fundamental band gap. However, in our case, electrical measurements show no significant variation in conductivity across doped and undoped films, suggesting that carrier concentration remains essentially unchanged. This rules out band gap renormalization as a contributing factor under the doping and growth conditions used in this study. The consistent red shift in optical absorption across all doped samples, regardless of dopant identity or valence, thus points to structural or defect-related perturbations introduced during growth, rather than true electronic doping. The consistently high hole conductivity across all films suggests that oxygen non-stoichiometry ($\delta \sim 0.15$) is the primary source of carriers, rather than extrinsic doping. Nonetheless, the introduction of dopant precursors may still influence the film's local structure or defect landscape[40]. For instance, dopants could introduce subtle lattice strain—either through ionic size mismatch or growth-induced stress—that perturbs the electronic structure. Strain effects on band gap are well-documented in other material systems; for example, reference [41] reports that in InGaAsP alloys, lattice mismatch induces measurable band gap shifts (~0.2% redshift per $\Delta x = 0.03$ in content). While our films do not show a clear trend correlating dopant ionic radius with optical redshift (e.g., Sc-doped films exhibit smaller shifts than others despite a larger ionic radius), the possibility of localized strain contributing to band gap narrowing cannot be ruled out.

Another possible mechanism is the dopant-induced modulation of oxygen incorporation during growth. Even slight variations in $\delta$ between samples—though not reflected in bulk conductivity—could alter the defect chemistry and affect the band edge through band tailing or the creation of localized states. Without precise measurements of hole concentration or oxygen content in each film, this remains speculative. However, the absence of visible-range absorption features or sub-gap states in the doped films indicates that the transparency is preserved and the primary absorption edge remains the dominant optical transition

In conclusion, extrinsic doping of $CuCrO_2$ via MOCVD yields only marginal changes to the material's properties, primarily manifesting as subtle band gap shifts attributable to many-body and disorder effects, rather than any dramatic changes in carrier density or phase composition. Doping does induce lattice strain/distortion (on a very small level) but the overall material remains dominated by its off-stoichiometric self-doping characteristics. Future work could explore alternative dopants or methods (e.g., post-annealing to drive in more dopants or modulation of oxygen content) to see if these trends hold, and to better separate the contributions of disorder vs. true band structure shifts in the optical phenomena.non-trivial and that the interplay between extrinsic and intrinsic defects is crucial.

**Conclusion**

This study investigated the effects of extrinsic doping on $CuCrO_2$ thin films grown by MOCVD using a uniform set of THD-based precursors for Cu, Cr, and all dopant species. Although dopants were introduced at nominal levels of 5 at.%, XPS and structural analysis revealed that actual incorporation was minimal—typically below 2 at.% and often below detection limits. Despite this, the films maintained high p-type conductivity (~35–70 S/cm), consistent with an intrinsic self-doping mechanism driven by oxygen excess ($\delta \sim 0.15$), rather than by effective substitutional doping. Optical characterization revealed a small but reproducible redshift in the band gap for all doped samples, yet this shift did not correlate with dopant type, valence, or nominal concentration. This behavior is more plausibly linked to strain effects, subtle structural disorder, or defect-mediated band tailing rather than to direct band structure modification through dopant incorporation. All films retained high optical transparency, with no evidence of sub-gap absorbers or plasmonic effects.

While doping was intended to modulate electronic properties, these results highlight a crucial point: dopant presence in the precursor does not guarantee incorporation into the lattice. Establishing a reliable structure–property relationship therefore requires direct quantification of film composition, rather than relying solely on precursor chemistry. Without this step, the interpretation of doping effects risks being misleading or incomplete. While numerous studies report high dopant concentrations, such values often reflect nominal precursor ratios or surface-sensitive measurements and may not represent true bulk substitution into the delafossite lattice. Without direct compositional quantification and phase analysis, apparent high doping levels can be misleading, especially in systems with low thermodynamic solubility.

Although direct bulk doping was largely unsuccessful, the presence of dopant precursors still subtly influenced film strain and optical response. This underscores an often-overlooked dimension in oxide doping studies: surface reactions, precursor volatility, and gas-phase interactions may significantly impact growth kinetics and defect formation—even in the absence of measurable substitutional incorporation. Future strategies should explicitly account for these effects, focusing not only on solubility limits but also on the broader chemical environment that governs film evolution during deposition.

## Acknowledgments

The authors would like to acknowledge the financial support by the Luxembourg National Research Fund- FNR (STEERMOB: Strain-Engineering the hole mobility in off-stoichiometric copper chromium delafossite - Project Number C19/MS/13577628).

## References

1. Chavan GT, Kim Y, Khokhar MQ, et al (2023) A Brief Review of Transparent Conducting Oxides (TCO): The Influence of Different Deposition Techniques on the Efficiency of Solar Cells. Nanomaterials 13:. https://doi.org/10.3390/nano13071226
2. Lunca-Popa P, Chemin JB, Adjeroud N, et al (2023) Study of Gallium-Doped Zinc Oxide Thin Films Processed by Atomic Layer Deposition and RF Magnetron Sputtering for Transparent Antenna Applications. ACS Omega 8:5475–5485. https://doi.org/10.1021/acsomega.2c06574
3. Woods-Robinson R, Morales-Masis M, Hautier G, Crovetto A (2024) From Design to Device: Challenges and Opportunities in Computational Discovery of $p$ -Type Transparent Conductors . PRX Energy 3:1. https://doi.org/10.1103/prxenergy.3.031001
4. Lyons JL, Janotti A, Van De Walle CG (2014) Effects of hole localization on limiting p-type conductivity in oxide and nitride semiconductors. J Appl Phys 115:012014. https://doi.org/10.1063/1.4838075
5. Varley JB, Miglio A, Ha VA, et al (2017) High-Throughput Design of Non-oxide p-Type Transparent Conducting Materials: Data Mining, Search Strategy, and Identification of Boron Phosphide. Chem Mater 29:2568–2573. https://doi.org/10.1021/acs.chemmater.6b04663
6. Brunin G, Rignanese G-M, Hautier G (2018) Are Small Polarons Always Detrimental to Transparent Conducting Oxides ? arxiv
7. Scanlon DO, Watson GW (2011) Understanding the p-type defect chemistry of CuCrO2. J Mater Chem 21:3655. https://doi.org/10.1039/c0jm03852k

8. Ingram BJ, Harder BJ, Hrabe NW, et al (2004) Transport and Defect Mechanisms in Cuprous Delafossites . 2 . CuScO2 and CuYO2. Chem Mater 16:5623–5629
9. Bai Z, Chen SC, Lin SS, et al (2021) Review in optoelectronic properties of p-type CuCrO2 transparent conductive films. Surfaces and Interfaces 22:100824. https://doi.org/10.1016/j.surfin.2020.100824
10. Afonso J, Greiner M, Lunca-Popa P, Raskin J-P (2023) Ultra-low-energy catalytic degradation of ozone by off-stoichiometric delafossite Cu- Cr-O for highly selective low-temperature solid-state O3 sensors. Mater Today Adv 18:100366. https://doi.org/10.1016/j.mtadv.2023.100366
11. Xu S, Zhao T, Kong L, et al (2021) Gas-solid interfacial charge transfer in volatile organic compound detection by CuCrO2nanoparticles. Nanotechnology 32:. https://doi.org/10.1088/1361-6528/abfa55
12. Kong L, Xu S, Liu H, et al (2022) Hybridized Ag-CuCrO 2 Nanostructured Composites for Enhanced Gas Sensing. Appl Nano Mater. https://doi.org/10.1021/acsanm.2c02529
13. Ahmadi M, Abrari M, Ghanaatshoar M (2021) An all-sputtered photovoltaic ultraviolet photodetector based on co-doped CuCrO2 and Al-doped ZnO heterojunction. Sci Rep 11:1–10. https://doi.org/10.1038/s41598-021-98273-5
14. Poienar M, Hardy V, Kundys B, et al (2012) Revisiting the properties of delafossite CuCrO 2: A single crystal study. J Solid State Chem 185:56–61. https://doi.org/10.1016/j.jssc.2011.10.047
15. Afonso J, Leturcq R, Lunca-Popa P, Lenoble D (2019) Transparent p-Cu0.66Cr1.33O2/n-ZnO heterojunction prepared in a five-step scalable process. J Mater Sci Mater Electron 30:1760–1766. https://doi.org/10.1007/s10854-018-0448-4
16. Kim J, Kendall O, Ren J, et al (2022) Highly Conductive and Visibly Transparent p-Type CuCrO2 Films by Ultrasonic Spray Pyrolysis. ACS Appl Mater Interfaces. https://doi.org/10.1021/acsami.1c24023
17. Zhang B, Thampy S, Dunlap-Shohl WA, et al (2019) Mg Doped CuCrO2 as Efficient Hole Transport Layers for Organic and Perovskite Solar Cells. Nanomaterials 9:1311. https://doi.org/10.3390/nano9091311
18. Willis J, Scanlon DO (2021) Latest directions in p-type transparent conductor design. J Mater Chem C 9:11995–12009. https://doi.org/10.1039/d1tc02547c
19. Bottiglieri L, Resende J, Weber M, et al (2021) Out of stoichiometry CuCrO 2 films as a promising p-type TCO for transparent electronics. Mater Adv 2:4721–4732. https://doi.org/10.1039/d1ma00156f
20. Lunca-Popa P, Crepelliere J, Nukala P, et al (2017) Invisible electronics: Metastable Cu-vacancies chain defects for highly conductive p-type transparent oxide. Appl Mater Today 9:184–191. https://doi.org/10.1016/j.apmt.2017.07.004
21. Norton E, Farrell L, Zhussupbekova A, et al (2018) Bending stability of Cu0.4CrO2 - A transparent p-type conducting oxide for large area flexible electronics. AIP Adv 8:4–11. https://doi.org/10.1063/1.5027038
22. Farrell L, Norton E, O'Dowd BJ, et al (2015) Spray pyrolysis growth of a high figure of merit, nano-crystalline, p -type transparent conducting material at low temperature. Appl Phys Lett 107:1–7. https://doi.org/10.1063/1.4927241
23. Lunca-Popa P, Botsoa J, Bahri M, et al (2020) Tuneable interplay between atomistic defects morphology and electrical properties of transparent p-type highly conductive off-stoichiometric Cu-Cr-O delafossite thin films. Sci Rep 10:1416. https://doi.org/10.1038/s41598-020-58312-z
24. Moreira M, Afonso J, Crepelliere J, et al (2022) A review on the p-type transparent Cu–Cr–O delafossite materials. J Mater Sci 57:3114–3142.

https://doi.org/10.1007/s10853-021-06815-z
25. Zhang KHL, Xi K, Blamire MG, Egdell RG (2016) P -type transparent conducting oxides. J Phys Condens Matter 28:383002. https://doi.org/10.1088/0953-8984/28/38/383002
26. Nagarajan R., Draeseke AD, Sleight AW, Tate J (2001) p-type conductivity in CuCr1-xMgxO2 films and powders. J Appl Phys 89:8022–8025. https://doi.org/10.1063/1.1372636
27. Ahmadi M, Asemi M, Ghanaatshoar M (2018) Mg and N co-doped CuCrO2 : A record breaking p-type TCO. Appl Phys Lett 113:242101. https://doi.org/10.1063/1.5051730
28. Lin SS, Shi Q, Dai MJ, et al (2020) The optoelectronic properties of p-type Cr-deficient Cu[Cr0.95-xMg0.05]O2 films deposited by reactive magnetron sputtering. Materials (Basel) 13:2376. https://doi.org/10.3390/ma13102376
29. Bottiglieri L, Nourdine A, Resende J, et al (2021) Optimized stoichiometry for CuCrO2 thin films as hole transparent layer in PBDD4T-2F: PC70BM organic solar cells. Nanomaterials 11:1–15. https://doi.org/10.3390/nano11082109
30. Moreira M, Crepelliere J, Leturcq R, et al (2024) Electrical properties of strained off-stoichiometric Cu – Cr – O delafossite. J Phys Condens Matter 36:215702
31. Lunca Popa P, Crêpellière J, Leturcq R, Lenoble D (2016) Electrical and optical properties of Cu – Cr – O thin fi lms fabricated by chemical vapour deposition. Thin Solid Films 612:194–201. https://doi.org/10.1016/j.tsf.2016.05.052
32. Moreira MAM (2024) The Influence of Induced Strain on Electric Properties of Nonstoichiometric Copper Chromium Oxide, PhD Thesis. University of Luxembourg
33. Ohtake A, Mano T, Sakuma Y (2020) Strain relaxation in InAs heteroepitaxy on lattice-mismatched substrates. Sci Rep 10:1–7. https://doi.org/10.1038/s41598-020-61527-9
34. Putkonen M, Nieminen M, Niinistö J, et al (2001) Surface-controlled deposition of Sc2O3 thin films by atomic layer epitaxy using β-diketonate and organometallic precursors. Chem Mater 13:4701–4707. https://doi.org/10.1021/cm011138z
35. Sartori A, El Habra N, Bolzan M, et al (2011) Stability study of a magnesium β-diketonate as precursor for chemical vapor deposition of MgO. Chem Mater 23:1113–1119. https://doi.org/10.1021/cm1020788
36. Guilmeau E, Poienar M, Kremer S, et al (2011) Mg substitution in CuCrO2 delafossite compounds. Solid State Commun 151:1798–1801. https://doi.org/10.1016/j.ssc.2011.08.023
37. Mugnier E, Barnabe A, Tailhades P (2006) Synthesis and characterization of CuFeO2+ delafossite powders. Solid State Ionics 177:607–612. https://doi.org/10.1016/j.ssi.2005.11.026
38. Miyasaka N, Doi Y, Hinatsu Y (2009) Synthesis and magnetic properties of ALnO2 (A=Cu or Ag; Ln=rare earths) with the delafossite structure. J Solid State Chem 182:2104–2110. https://doi.org/10.1016/j.jssc.2009.05.035
39. Crepelliere J, Moreira M, Lunca-Popa P, Leturcq R (2025) On the charge transport models in high intrinsic defect doped transparent and conducting p-type Cu – Cr – O delafossite. J Phys D Appl Phys 58:015310
40. Benreguia N, Rekhila G, Younes A, et al (2024) Optical and electrical properties of the delafossite CuCrO2 synthesized by co-precipitation. Inorg Chem Commun 162:. https://doi.org/10.1016/j.inoche.2024.112154
41. Kim MR, Kim CH, Han BH (1998) Band-gap renormalization and strain effects in semiconductor quantum wells. Phys B Condens Matter 245:45–51. https://doi.org/10.1016/s0921-4526(97)00450-x